\documentclass[reprint, amsmath,amssymb,aps,pre,superscriptaddress,longbibliography,notitlepage]{revtex4-2}
\usepackage[normalem]{ulem}
\usepackage{url}
\usepackage{hyperref}
\usepackage{xcolor}
\usepackage{graphicx}
\usepackage{dcolumn}
\usepackage{bm}
\usepackage{url}
\usepackage{soul}
\usepackage{hyperref}
\newcommand{\fsdel}[1]{}

\newcommand{\rem}[1]{}

\newcommand{\hhr}{\hat{r}}
\newcommand{\hht}{\hat{t}}
\newcommand{\hbr}{\hat{\mathbf{r}}}
\newcommand{\hhl}{\hat{\lambda}}
\newcommand{\hhs}{\hat{s}}
\newcommand{\hhF}{\hat{\mathcal F}}
\newcommand{\hhFeff}{\hat{\mathcal F}^{\text{eff}}_t}
\newcommand{\tphi}{\tilde{\boldsymbol{\phi}}}
\newcommand{\rO}{\rho_{_0}}
\newcommand{\lO}{\ell_{_0}}
\newcommand{\tO}{t_{_0}}
\newcommand{\qO}{q_{_0}}
\newcommand{\rone}{\rho_{_1}}
\newcommand{\rtwo}{\rho_{_2}}
\newcommand{\ldinf}{l_\text{e}}

\newcommand{\bhhl}{\hat{\boldsymbol{\lambda}}}

\begin{document}
\title{Active Transport as a Mechanism of Microphase Selection in Biomolecular Condensates}

\author{Le Qiao}
\email{le.qiao@uni-mainz.de}
\affiliation{Institute of Physics, Johannes Gutenberg University Mainz, D55099 Mainz, Germany}
\author{Peter Gispert}
\affiliation{Institute of Physics, Johannes Gutenberg University Mainz, D55099 Mainz, Germany}
\author{Lukas S. Stelzl}
\affiliation{Institute of Molecular Physiology, Johannes Gutenberg University Mainz, D55099 Mainz, Germany}
\affiliation{Institute of Molecular Biology (IMB), Mainz, Germany}
\author{Friederike Schmid}
 \email{friederike.schmid@uni-mainz.de}
\affiliation{Institute of Physics, Johannes Gutenberg University Mainz, D55099 Mainz, Germany}
\date{\today}
\begin{abstract}
The size and organization of biomolecular condensates formed by liquid-liquid
phase separation (LLPS) are set by multiple cellular mechanisms that are not yet
fully understood. Here we identify a transport-driven mechanism: stochastic
binding of phase-separating proteins to cytoskeletal motor proteins, followed by
active redistribution along filament networks, generates an effective long-range
repulsion that arrests coarsening and selects a finite condensate size. A
minimal diffusion-transport model, analyzed by linear stability theory and
three-dimensional simulations, reveals a transition from macroscopic to
microphase separation at remarkably low binding/release fractions, corresponding
to minute motor-bound populations. Tuning motor binding rates $b$ or
transport velocities enables sublinear control of condensate sizes
($L \sim b^{-1/4}$) from a few hundred nanometres up to the micron scale. The selected length scale is robust to the intrinsic shot noise of the
binding--release reactions. In anisotropic cytoskeletal environments, transport asymmetry drives morphological transitions from spherical to cylindrical condensates, independently of the thermodynamic parameters. This mechanism provides a versatile, spatiotemporally
programmable route to condensate organization and informs the design of
synthetic active emulsions with tunable architectures.
\end{abstract}

\maketitle

\section{Introduction}
Eukaryotic cells compartmentalize their biochemical activities through both membrane-bound organelles (e.g., mitochondria, lysosomes) and membraneless assemblies (e.g., nucleoli, centrosomes, stress granules). Many of the latter arise via liquid–liquid phase separation (LLPS), forming dynamic, liquid-like condensates composed of proteins and RNA~\cite{brangwynneGermlineGranulesAre2009a,albertiConsiderationsChallengesStudying2019,brangwynnePhaseTransitionsSize2013,albertiBiomolecularCondensatesNexus2021,hymanLiquidLiquidPhaseSeparation2014}. These condensates enable reversible, spatially confined reactions without enclosing membranes. Conversely, dysregulation of condensate dynamics or aging can trigger irreversible amyloid fibril formation~\cite{strooCellularRegulationAmyloid2017,linsenmeierInterfaceCondensatesHnRNPA12023}, a hallmark of neurodegenerative diseases such as Alzheimer's, Parkinson's, and ALS. 

Under physiological conditions, dilute and condensed phases coexist near equilibrium~\cite{shinLiquidPhaseCondensation2017, vecchiProteomewideObservationPhenomenon2020, alertFormationMetastablePhases2016}, exhibiting slow coarsening dynamics governed by fusion and Ostwald ripening~\cite{brayTheoryPhaseorderingKinetics2002}. Left unchecked, this slow relaxation would ultimately yield a single macroscopic droplet. To arrest coarsening and stabilize finite‐sized condensates, cells exploit various nonequilibrium mechanisms\cite{weberPhysicsActiveEmulsions2019}. Chemical reaction–diffusion cycles, such as kinase–phosphatase or autocatalytic conversions, can generate interfacial fluxes that balance diffusive loss and select a finite length scale~\cite{glotzerChemicallyControlledPattern1994,glotzerReactionControlledMorphologyPhaseSeparating1995a,christensenPhaseSegregationDynamics1996,caratiChemicalFreezingPhase1997,zwickerSuppressionOstwaldRipening2015,wurtzChemicalReactionControlledPhaseSeparated2018,hondeleDEADboxATPasesAre2019,kirschbaumControllingBiomolecularCondensates2021,schedeModelOrganizationRegulation2023,ziethenNucleationChemicallyActive2023,hafnerReactiondrivenAssemblyControlling2023,zippoMolecularSimulationsEnzymatic2025,harmonMolecularAssemblyLines2022,braunsWavelengthSelectionInterrupted2021}. Diffusiophoretic forces arising from ion gradients~\cite{doanDiffusiophoresisPromotesPhase2024}, macromolecular crowding~\cite{friesChemicallyActiveDroplets2025}, protein oligomerization~\cite{rossettoBindingDimerizationControl2025,litschelMembraneinduced2DPhase2024}, long‐range electrostatic repulsion~\cite{luoTheoryCondensateSize2025}, Pickering emulsion mechanism\cite{folkmannRegulationBiomolecularCondensates2021b} or elastic confinement by the cytoskeleton~\cite{weiModelingElasticallyMediated2020,liuLiquidLiquidPhase2023,styleLiquidLiquidPhaseSeparation2018,yokoyamaMolecularSimulationsPhase2026} can also impede droplet growth. More recently, coupling a phase-separating species to a self-straining cytoskeletal network has been shown to arrest coarsening mechanically~\cite{bodini--lefrancArrestedCoarseningOscillations2025}, though the broader role of stochastic motor-mediated transport in regulating condensate sizes remains largely unexplored. 

Many proteins and RNA molecules that are constituents of phase-separated
condensates are transported by motor proteins to defined subcellular
locations~\cite{Kiebler:NatRevNeuroScience:2024,appert-rollandIntracellularTransportDriven2015}.
The microtubule-associated motor protein dynein co-localizes with stress
granules, which help to protect cells during acute
stresses~\cite{tsaiDyneinMotorContributes2009}. Inhibiting dynein reduces
stress granule formation and delays their disassembly once the stress
subsides~\cite{tsaiDyneinMotorContributes2009}. Knockdown of dynein likewise
inhibits stress granule formation, whereas knockdown of kinesin delays their
dissolution~\cite{Loschi:JCellSci:2009}. A motor-based transport system for RNA
protein complexes and protein condensates has been realized in cells and this
could be a starting point for understanding how transport affects phase
behavior~\cite{cochardCondensateFunctionalizationMicrotubule2023}.

\begin{figure}
    \centering
\includegraphics[width=.96\linewidth]{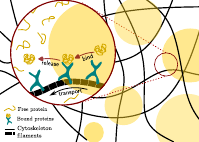}
\caption{Schematic cartoon of the proposed active transport mechanism to control droplet sizes. Magnified view shows: Phase-separating 
proteins (yellow) exchange between a free diffusive state and a motor-bound 
transported state at rates $b$ and $s$. Motor-mediated transport along 
cytoskeletal filaments (black) spatially redistributes bound proteins, creating 
an effective long-range repulsion that arrests coarsening at finite length scale.
}
\label{fig:1new}
\end{figure}

Here, we show that motor-mediated transport along cytoskeletal filaments is sufficient on its own to arrest condensate coarsening and stabilize finite domain sizes. In contrast to chemical reaction-diffusion mechanisms, motor binding does not alter the chemical properties of proteins and their molecular interactions. Starting from a minimal diffusion–transport model, we derive an analytical criterion for the onset of microphase separation and obtain a scaling law that directly connects condensate dimensions to motor binding rates and transport velocities. Three-dimensional simulations confirm these predictions across a wide range of noise levels and quench depths, and further reveal that cytoskeletal anisotropy extends this control from size to shape, driving transitions between spherical, cylindrical, and lamellar morphologies.
A schematic cartoon is shown in Fig.~\ref{fig:1new}.

\section{Model and Physical Intuition}
Our model is inspired by the Foret model for membrane rafts, which shows how actively driven lipid turnover can confine domain sizes in phase separating lipid membranes~\cite{foretSimpleMechanismRaft2005}. In that model, lipid molecules are stochastically removed from or inserted into the membranes at fixed rates. This process turns out to be mathematically equivalent to introducing an effective long-range repulsion between like species~\cite{garckecoupled2016, schmidPhysicalMechanismsMicro2017}, which arrests coarsening and stabilizes microdomains. Here, we develop a three-dimensional bulk analogue of this mechanism. Proteins interconvert between a free state, $\rone$, and a transported state, $\rtwo$, in which they are actively carried along cytoskeletal filaments by motor proteins, with binding and release rates $b$ and $s$. In this context, active transport plays a role analogous to stochastic removal and insertion. Motor-driven redistribution actively homogenizes the bound protein fraction, again generating effective long-range interactions. The coupled dynamical equations are
\begin{align}
  \partial_t \rone
  &= M \,\nabla^2\!\Bigl(\frac{\delta {\mathcal  F}}{\delta\rone}\Bigr)
   -b\,\rone + s\,\rtwo
   + \eta_e,
  \label{eq:rho1}\\
  \partial_t \rtwo
  &= \nabla \,\boldsymbol{\Lambda} \,\nabla\rtwo 
   +b\,\rone - s\,\rtwo
   - \eta_e.
  \label{eq:rho2}
\end{align} 

Equation~\ref{eq:rho1} describes the diffusive relaxation of free proteins driven by the chemical potential gradient $\nabla\delta\mathcal{F}/\delta\rho_1$, with Onsager mobility $M$, together with binding/unbinding exchange with rates $b$ and $s$; the accompanying chemical noise $\eta_e$ is modeled via Langevin dynamics~\cite{gillespieChemicalLangevinEquation2000}. Equation~\ref{eq:rho2} captures the active transport of motor-bound proteins. We consider bidirectional random transport along a disordered network of filaments\cite{mullerBidirectionalTransportMolecular2010,joshiEmergentSpatiotemporalOrganization2024}, which can be coarse-grained as rapid, anisotropic diffusion with an effective diffusion tensor $\boldsymbol{\Lambda}$\cite{caspiEnhancedDiffusionActive2000,brangwynneIntracellularTransportActive2009}. In an isotropic cytoskeletal network, we have $\boldsymbol{\Lambda} = \lambda \boldsymbol{I}$, whereas filament alignment leads to anisotropic transport described by a uniaxial diffusion tensor with eigenvalues $\lambda_\parallel > \lambda_\perp$, reflecting enhanced mobility along filament bundles.\cite{fredricksonAnalyticalSolutionModel1995,
joshiEmergentSpatiotemporalOrganization2024,brangwynneIntracellularTransportActive2009} For analytical tractability, we assume isotropy in the scaling and linear stability analyses,  and explore the effects of anisotropic transport by numerical simulations. The free energy functional, ${\mathcal F}[\rho]$, depends on the total density
$\rho = \rone + \rtwo$ via
\begin{equation} 
 \frac{\mathcal F[\rho]}{\rO k_B T} = 
    \int_V \!\!\! \mathrm{d}^3r\,
    \Bigl[
      \frac{\tau_2}{2 \rO^2}(\rho-\rho_c)^2
      +\frac{\tau_4}{4 \rO^4}(\rho-\rho_c)^4
      +\frac{g}{2 \rO^2}|\nabla\rho|^2
    \Bigr]
    \label{eq:F}
\end{equation}
where $\tau_2$ and $\tau_4 $ control bulk thermodynamics, $\rO=\bar\rho$ and $\rho_c$ are the averaged total density and the critical density, $V$ is the system volume, and $g$ sets the interfacial stiffness. The noise accounts for the discrete nature of stochastic binding/unbinding events and is modeled by chemical Langevin dynamics~\cite{gillespieChemicalLangevinEquation2000} (see also SI, Section I.A)
\begin{equation}
\eta_e (\mathbf{r},t) 
= \sqrt{b \rone(\mathbf{r})} \: \zeta_1(\mathbf{r},t)
- \sqrt{s \rtwo(\mathbf{r})} \: \zeta_2(\mathbf{r},t),
\end{equation}
where $\zeta_i(\mathbf{r},t)$ are uncorrelated white noise fields
(It\^o interpretation~\cite[Sec.~2.3, p.~40]{paulbaschnagel_book})
with $\langle \zeta_i \rangle \equiv 0$ and
$\langle \zeta_i (\mathbf{r},t) \zeta_j (\mathbf{r}',t') \rangle 
= \delta_{ij} \, \delta(\mathbf{r}-\mathbf{r}') \, \delta(t-t')$.\\

Simple scaling arguments can provide intuition on the characteristic domain size. Balancing binding kinetics (binding-release rate, $b$ and $s$), free protein diffusion ($M k_BT /\rO$), and transport ($\lambda$) yields two distinct regimes: (I) In the \emph{exchange-limited} regime, transport processes are fast enough to be quasi-instantaneous. Molecules are redistributed rapidly and become available to participate in LLPS elsewhere. Thus domain growth is limited by the competition between binding kinetics and structural relaxation, which is set by the Cahn–Hilliard $\nabla^4$ dissipation. A fluctuation of size $L$ relaxes on a time scale 
$t_{M}\sim (L^4 \rO)/(M g k_B T)$. 
Matching the binding time, $t_{b}\sim 1/b$, gives 
\begin{equation}
L\sim \ldinf := 
    \Bigl(\frac{M\,k_B T\, g}{\rO\, b}\Bigl)^{1/4}.
    \label{eq:ldinf}
\end{equation}
(II) In the \emph{diffusive} regime, transport is too slow to ensure complete redistribution of proteins. If free proteins bind and get released within one droplet, the droplet growth is not arrested. Bound molecules travel a characteristic distance $\Delta l_\text{a} = \sqrt{\lambda/s}$.
The transition between the two regimes is thus expected at $\ldinf \sim \Delta l_\text{a}$ or 
\begin{equation}
    \kappa := \frac{\ldinf}{\Delta l_\text{a}} =  \Bigl(\frac{M\,k_BT\, g \, s^2}{\rO\, b\, \lambda^2}\Bigl)^{1/4} \approx 1.
    \label{eq:kappa}
\end{equation}

For a more quantitative analysis, we first cast the equations in dimensionless form, using scaled lengths $\hhr = r/\lO$ and times $\hht = t/\tO$ with $\lO =\sqrt{g}$ and $\tO = \ (g \rO)/(M k_B T)$ and defining the order parameters, $\phi_1=(\rone-\bar{\rho}_1)/\rO$ and $\phi_2=(\rtwo-\bar{\rho}_{_2})/\rO$ with $\bar{\rho}_{_1}=\rO \, s/(b+s)$ and $\bar{\rho}_{_2}=\rO \, b/(b+s)$. In physical terms, $\lO$ is of the order of the size of a protein, $\lO^2/\tO$ is roughly the molecular diffusion constant of free proteins, and $\bar{\rho}_{_{1,2}}$ correspond to the mean concentrations of unbound/bound proteins in the homogeneous fluid.

The resulting rescaled dynamic equations Eqs.\ \ref{eq:phi1}--\ref{eq:phi2}
(see Appendix \ref{sec:appendix}) reveal that the behavior of the system is governed by four dimensionless quantities: The binding/release ratio $e = b/s$ setting the partitioning of molecules between the two states, the dynamic asymmetry $\hhl = (\lambda \rO)/(M k_B T)$ giving the ratio of random transport in the bound state and free diffusion in the unbound state, the scaled release
rate $\hhs =(s g \rO)/(M k_B T)$, and the noise level parameter $C_0= g^{-3/4}/\sqrt{\rO}$ (see SI). In these units, the characteristic quantities introduced above are given by $\ldinf = (\hhs e)^ {-1/4} \: \lO$ and $\kappa = (e \, \hhl^2/\hhs)^{-1/4}$.\\
\section{Linear Stability Analysis and Phase Diagram}
\label{sec:LSA}
\begin{figure}[ht]
    \centering
\includegraphics[width=0.96\linewidth]{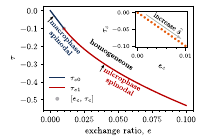}
\caption{Stability diagram in the $(e,\tau)$ plane. The blue branch $\tau_{c0}(e)$
is the spinodal with respect to macrophase separation, the red branch
$\tau_{c1}(e)$ that with respect to microphase separation; they meet at the
critical point $(e_c,\tau_c)$. The labels refer to the character of the
instability at onset, not to a division of the two-phase region. Parameters are
$\hhs=1$ and $\hhl=10$. Inset: shift of the critical point as the scaled release
rate $\hhs$ increases from $0.01$ to $1$.}
    \label{fig:2new}
\end{figure}

Next we examine the stability of the homogeneous state ($\phi_i \equiv$ const.) against small perturbations. It is instructive to consider first a simplified model, in which we assume that the fraction of transported molecules is small and replace $\mathcal F[\rho]$ by $\mathcal F[\rone]$ in the free energy Eq.\ \ref{eq:F}. This approximation allows us to map the noise-free part of our model onto an equilibrium model governed by purely diffusive relaxation. The key step is to represent the source and sink terms in Eqs.\ \ref{eq:rho1}--\ref{eq:rho2} as equivalent long-range Coulomb interactions. Similar transformations have been employed to describe arrested phase
separation in the Foret model of lipid rafts~\cite{garckecoupled2016,schmidPhysicalMechanismsMicro2017} and pattern formation in certain reactive binary mixtures~\cite{glotzerChemicallyControlledPattern1994,
glotzerReactionControlledMorphologyPhaseSeparating1995a,
liNonequilibriumPhaseSeparation2020}. We first symmetrize the exchange terms by rescaling the
order parameter, $\tphi = (\phi_1,\phi_2/\sqrt{e})$. 
After some mathematical manipulations detailed in the SI, the dynamical equations can be rewritten as
\begin{equation}
\partial_{\hht} \tphi = 
  \hat{\nabla}^2 \frac{\delta{\hhF_\text{eff}}}{\delta \tphi}
  +  \hat{\eta}_e(\hbr,t) \: {1\choose -1/\sqrt{e}}
  \label{eq:dynamics_simple}
\end{equation}
in terms of a virtual effective free energy functional
\begin{align}
\hhFeff &= 
\frac{1}{2\hat{V}} \sum_{\mathbf{q} \neq 0} \tphi^\dagger_\mathbf{q}\, 
  \mathbf{B}_\mathbf{q}\, \tphi_\mathbf{q} 
 +   \int_{\hat{V}} \!\! \text{d}^3\hhr\,
   \Bigl[
      -  \tau_4 \phi_c \: \tilde{\phi}_1^3 
      + \frac{\tau_4 }{4}\tilde{\phi}_1^4
    \Bigr]  
  \\ 
\text{with} &\quad  \mathbf{B}_\mathbf{q} = \begin{pmatrix}
\tau + q^2 + e \hhs/q^2 & -\sqrt{e} \hhs/q^2 \\
-\sqrt{e} \hhs/q^2 & \hhl + \hhs/q^2
\end{pmatrix}.
\label{eq:feff}
\end{align}
\noindent
Here $\tau=\tau_2+3 \tau_4 \phi_c^2$ with $\phi_c := \rho_c/\rO -1$, $\hat{V}=V/\lO^3$, and $\hat{\eta}_e$ is an uncorrelated noise field with mean zero and variance 
$C_0^2 \hhs e(\tilde{\phi}_1 + \tilde{\phi}_2/\sqrt{e} + 2/(1+e))$ (see SI).
In the absence of noise, the homogeneous solution of the dynamical equation
\ref{eq:dynamics_simple}, $\tphi \equiv 0$, is (meta)stable whenever $\mathbf{B}(q)$ is positive definite, and becomes unstable when $\det(\mathbf{B}(q)) = 0$. This condition yields two branches of the spinodal, $\tau_{c0}(e)= -e\,\hhl$ and $\tau_{c1}(e)=-2\sqrt{e\,\hhs} + \hhs/\hhl$, which meet at a critical exchange ratio $e_c = \hhs/\hhl^2$ (see Fig.~\ref{fig:2new}). For $e < e_c$, the unstable mode occurs at $q \to 0$, corresponding
to macroscopic (bulk) phase separation. For $e > e_c$, the instability appears at a finite wavenumber $q=\qO$, marking the onset of a microphase regime with characteristic domain size
$L \sim 2 \pi/ \qO$.
The domain size $L$ diverges at the critical point 
$\bigl(e_c,\,\tau_c\bigr) = \bigl(\hhs/\hhl^2,\,-\hhs/\hhl\bigr)$, where the most unstable mode shifts from $q=0$ to $q>0$. 

At the spinodal $\tau=\tau_{c1}$, the characteristic wave number is given by 
\begin{equation}
     \qO^2=\ell_{_0}^{-2}\left(\sqrt{e\hhs}-\hhs/\hhl\right).
\end{equation}
Expressing the domain size in terms of the quantities $\ldinf$ and $\kappa$ introduced 
in Eqs.\ \ref{eq:ldinf}--\ref{eq:kappa}, one gets
\begin{equation}
L = 2\pi\,\ldinf (1-\kappa^2)^{-1/2}.
\end{equation}
The domain size diverges for $\kappa \to 1$ and approaches $2\pi\,\ldinf$ for $\kappa \to 0$, consistent with our earlier scaling arguments. Furthermore, noting that $\kappa = (e_c/e)^{1/4}$, one can extract the power law $L \sim (2\pi\,\ldinf \sqrt{e_c/2}) \: (e-e_c)^{-1/2}$ describing how the domain size diverges upon approaching $e_c$.

The structure of the functional $\hhFeff$ in Eq.\ \ref{eq:dynamics_simple} closely
resembles the Ohta-Kawasaki-Ito functional for homopolymer/copolymer blends~\cite{ohtadynamics1995}. Such systems exhibit both Ising-type macrophase separation and Landau-Brazovskii-type~\cite{brazovskiiphase1975} microphase separation, separated by a multicritical Lifshitz point where the characteristic domain size diverges -- consistent with our stability analysis. It is important to emphasize, however, that the nonequilibrium model defined by Eq.\ \ref{eq:dynamics_simple} is {\em not equivalent} to an equilibrium model with free energy $\hhFeff$. The stochastic term violates the fluctuation dissipation relation, meaning that even in the limit $C_0 \to 0$, globally minimizing $\hhFeff$ will not necessarily yield the correct steady-state phase diagram, particularly in the presence of competing metastable states.

In the full model, $\mathcal F[\rho]$ with $\rho=\rho_1+\rho_2$, the
exchange terms cannot be symmetrized, therefore the mapping onto an  equilibrium system with an effective free energy  
$\hhFeff$ of Ohta-Kawasaki type is not possible. Nevertheless, we can still perform a stability analysis. The stability boundary is obtained directly from the condition that the linearized dynamical matrix has a vanishing eigenvalue
(see SI). The remaining algebra is unchanged, and yields the spinodal
\begin{equation}
  \tau_{c1}^{\rm F}(e) = -2\sqrt{e\hhs} + \frac{\hhs\,(1+e)}{\hhl},
\end{equation}
with the corresponding wavenumber
$( \qO^{\rm F})^2=\ell_{_0}^{-2}\bigl[\sqrt{e\hhs}-\hhs(1+e)/\hhl\bigr]$ and domain size $L^{\rm F}=2\pi/ \qO^{\rm F}$. Both
expressions reduce to those of the simplified model if one replaces $\hhs(1+e)\to\hhs$, thus the entire effect of retaining $\phi_2$ is this single substitution. The spinodal
of the full model is shifted upwards by
\begin{equation}
  \delta\tau_{c1}(e) \;=\; \tau_{c1}^{\rm F}(e)-\tau_{c1}(e)
  \;=\; \frac{\hhs\,e}{\hhl} \;>\;0
\end{equation}
Since the transported proteins also contribute to the driving force for
demixing, phase separation sets in slightly earlier. The shift vanishes
as $e\to0$ and grows linearly with $e$; for the parameters used here it amounts
to $\delta\tau_{c1}=2\times10^{-3}$ at $e=0.02$ and $5\times10^{-3}$ at $e=0.05$,
and the corresponding domain size is larger by
$\delta L/L\simeq \hhs e/\bigl[2 \hhl \, (\qO \lO)^2\bigr]\approx2\%$ across the range studied in the numerical simulations of the next section. The approximation underlying $\hhFeff$ is therefore very good in the parameter regions considered here, while becoming less so on approaching the transition point, $e=e_c$, where $ \qO \to 0$.

\section{Three Dimensional Simulations}

\label{sec:V}
To test our analytical predictions and explore the nonlinear regime, we performed direct numerical simulations. We solved the rescaled versions of Eqs.\ \ref{eq:phi1}--\ref{eq:phi2} (see Appendix A), fully taking into account the $\rtwo$ contribution to the free energy ${\cal F}$ in Eq.\ \ref{eq:F}. For comparison, we also simulated the approximate model used for the stability analysis; these results are presented in the SI.

\subsection{Microphase Arrest and Robustness}
\begin{figure*}[!ht]
  \centering
\includegraphics[width=\textwidth]{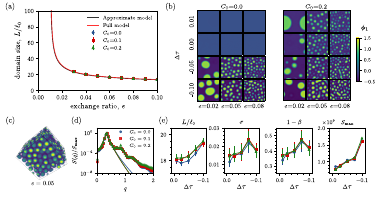}
  \caption{Transport‐driven domain selection in 3D simulations.  (a) Steady-state domain size $L/\lO$ vs exchange ratio $e$:  theory (black solid line for approximate model and red solid line for full model) and simulations with different noise strength $C_0=0,\,0.1,\,0.2$ for $\Delta \tau=: \tau-\tau_{c1}^{\rm F}(e)=-0.01$ (symbols). For $e = 0.02$, the domain sizes keep growing with a $\hht^{1/3}$ power law at the end of the simulation (movies and additional data see SI Fig.~S3). (b) Midplane $\phi_1$ slices for $e=0.02,\,0.05,\,0.08$ with and without noise at time $\hht=5 \times 10^4$. Rows correspond to four values of the quench parameter $\Delta\tau=0.01,\, 0.00,\, -0.05,\,-0.10$ where $\Delta\tau=0$ corresponds to the spinodal. (c) 3D snapshot for $e=0.05$ at time $\hht=5 \times 10^4$. (d) Angle‐averaged structure factor $S(q)$ for $e=0.05$ at $\Delta\tau=-0.01$ and noise strengths $C_0=0,\,0.1,\, 0.2$, plotted in semi-log representation. Solid lines are fits to the empirical form $S(q)\sim\exp\{1 - [1 + ((q- \qO)/\sigma)^2]^{1-\beta}\}$. (e) Fitted $L$, peak width $\sigma$, exponent $1-\beta$, and peak amplitude $S_{\max}$ vs quench depth $\Delta\tau$ for $e=0.05$. Parameters are $\hhs=1$, $\hhl=10$, $\phi_c=0.4$, $\tau_4 = 0.4$. }
  \label{fig:3}
\end{figure*}
Figure~\ref{fig:3}(a) shows that the finite-$q$ instability quantitatively predicts the steady‐state droplet size $L$. For large $e$, coarsening arrests and $L$ remains finite, while for small $e$, droplets continue growing, following the characteristic $\hht^{1/3}$ growth law during the whole simulation time of up to $\hht=5 \times 10^4$ (data in SI), which is characteristic of macrophase separation. The transition from microphase to macrophase separation predicted by the linear
theory is reproduced by the simulations, but the boundary lies at a slightly higher $e$ than the analytic $e_c$.
The offset is present in the simplified-model simulations as well (see SI). This observation is consistent with the phase behavior of the corresponding effective equilibrium system, where Lifshitz points are generally known to be unstable in three dimensions, and preempted by a multiphase equilibrium point in block copolymer/homopolymer mixtures\cite{vorselaars2020instability}.

Panel~(b) presents midplane snapshots of $\phi_1$ for $\phi_c =0.4$, three values of $e$, and
four quench depths $\Delta\tau:=\tau-\tau_{c1}^{\rm F}(e)$, comparing simulations
without and with noise ($C_0=0$ and $C_0=0.2$). Without noise, no structure forms
at $\Delta\tau=0$ or above, as expected: at the spinodal, the growth rate of the
critical mode vanishes, therefore an infinitesimal perturbation is not amplified within
the simulated time. Upon adding noise, droplets appear both at the spinodal and above
it at $\Delta\tau=0.01$, where the homogeneous state is linearly stable. The morphology of structures already formed at deeper quenches, however, is largely unaffected by the noise. The noise-induced structure formation can again be rationalized based on the phase diagram of the corresponding effective equilibrium system: At $\phi_c=0.4$, corresponding to the mean density $\rO = 0.71 \rho_c$, the order-disorder transition is first order; the spinodal lies inside the ordered region. Therefore, noise can induce microphase separation. Furthermore, the system exhibits hysteresis: Once established, the domains do not disappear even if the noise is turned off, see SI Section\ V.

To quantify noise effects more systematically, we analyze the structure factor in panels~(d) and (e). Panel~(d) displays the normalized structure factor $S(q)/S_{\max}$ for $e=0.05$, revealing a sharp primary peak at $\qO$ and secondary oscillations at larger $q$. These post-peak oscillations signal short-range translational order, indicating that the droplets locally adopt a quasi-regular arrangement reminiscent of a soft crystal or microphase. Introducing noise predominantly increases the spectral weight at large $q$, effectively lifting the high $q$ tail of $S(q)$. As shown in the SI, this follows from the noise adding a $q^{-4}$ background of amplitude
$\propto C_0^2$, which dominates once the deterministic tail has decayed. Finally, panel~(e) shows that deeper quenches ($\Delta\tau<0$) modestly increase $L$ and drive the exponent $(1-\beta)\to0.5$, indicating a crossover toward large-$q$ exponential decay in $S(q)$. In this regime, noise slightly enlarges $L$, consistent with enhanced interfacial fluctuations. These trends are nearly independent of noise amplitude, underscoring the robustness of the transport‐driven length‐scale selection.\\ 

\subsection{Anisotropic Transport}
While intracellular transport may appear disordered, cytoskeletal organization often leads to anisotropy in the effective diffusion of bound proteins. Cytoskeletal filament networks, even when not forming ordered bundles, can exhibit preferred orientations due to centrosomal organization or cell geometry. To investigate how such transport anisotropy affects condensate formation, we replace the scalar diffusion $\hhl$ with an anisotropic tensor $\boldsymbol{\Lambda} = \text{diag}(\hhl_\parallel, \hhl_\perp, \hhl_\perp)$ in numerical simulations, representing enhanced transport along a preferred direction. 

\begin{figure}
    \centering
\includegraphics[width=0.96\linewidth]{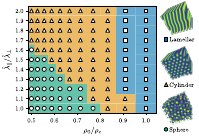}
\caption{Morphology phase diagram for anisotropic transport. Steady-state morphologies in $(\rO/\rho_c, \hhl_\parallel/\hhl_\perp)$-space: Lamellar (squares), cylinders (triangles), spheres (circles). Increasing transport anisotropy drives sphere-to-cylinder transitions. Parameters: $\hat{s} = 1$, $\tau_4 = 0.2$, $\hhl_\perp = 10$, 
$e = 0.06$, $\Delta\tau = -0.01$, $128^3$ grid.}\label{fig:ani_phase}
\end{figure}

Transport anisotropy does not disrupt microphase separation, but can have a profound effect on the morphologies of microphase separated domains. Figure~\ref{fig:ani_phase} maps steady-state morphologies across the parameter space of scaled average protein concentration, $\rO/\rho_c=1/(1+\phi_c)$, and transport anisotropy, $\hhl_\parallel/\hhl_\perp$. The interaction parameters are chosen such that $\Delta \tau = -0.01$, slightly below the (micro)phase separation temperature in the corresponding isotropic system. All simulations were initialized from steady-state spherical droplets (obtained at $\hhl_\parallel = \hhl_\perp$ and $\phi_c=0.4$) and evolved up to the time $\hht = 5 \times 10^4$. We classify final morphologies as: \textbf{(i)} spheres if droplets remain approximately round; \textbf{(ii)} cylinders if they elongate into rods aligned with the fast transport axis; or \textbf{(iii)} lamellar if they reorganize into layered structures. Phase boundaries mark where spherical droplets become visibly elongated.

Already in the isotropic systems, cylindrical and lamellar condensate morphologies appear at near-critical protein concentrations ($\rO/\rho_c \sim 1$). This is to be expected, since such morphological transitions are generally observed in Landau-Brazovskii-type models~\cite{brazovskiiphase1975, ohtaequilibrium1986}. Transport anisotropy shifts these transitions to  biologically more relevant regimes of smaller protein concentrations. For example, we observe a sphere-to-cylinder transition  for $\rO/\rho_c \sim 0.7$ at modest $\hhl_\parallel/\hhl_\perp \gtrsim 1.1$, shifting to higher anisotropy values as $\rO/\rho_c$ decreases further. The transition mechanism involves a transport-driven morphological instability. Spherical droplets experience enhanced flux along the parallel direction, which amplifies elongation fluctuations aligned with the fast axis. The resulting cylinders minimize interfacial area in slow-transport directions while extending along the fast axis where material redistributes more efficiently. The  parameter $\rO/\rho_c$, controlled by the total volume fraction of condensates, modulates these transitions. Lower $\rO/\rho_c$ corresponds to greater compositional asymmetry and favors compact spherical domains over thin cylindrical domains, analogous to minority-block spheres in copolymers.\\

\section{Biological and Experimental Implications}

To translate our results into physical units, we identify the molecular
scales $\lO = 1\,\mathrm{nm}$ and $\tO = 1\,\mu\mathrm{s}$, corresponding
to a free-protein diffusion constant
$\lO^2/\tO = 10^{-12}\,\mathrm{m}^2/\mathrm{s}$, and adopt a typical cargo
release rate of $0.1\,\mathrm{s}^{-1}$~\cite{gilliesCargoAdaptorIdentity2025},
giving $\hhs = 10^{-7}$. The dynamic asymmetry $\hhl$ compares the
effective transport of motor-bound proteins with the diffusion of free ones.
Motor activity in cells generates diffusive-like motion of substantially larger
amplitude than thermal diffusion~\cite{brangwynneIntracellularTransportActive2009,caspiEnhancedDiffusionActive2000}, thus the
biologically relevant regime is $\hhl > 1$; the precise value depends on motor type and processivity. 

\begin{figure}
    \centering
\includegraphics[width=0.96\linewidth]{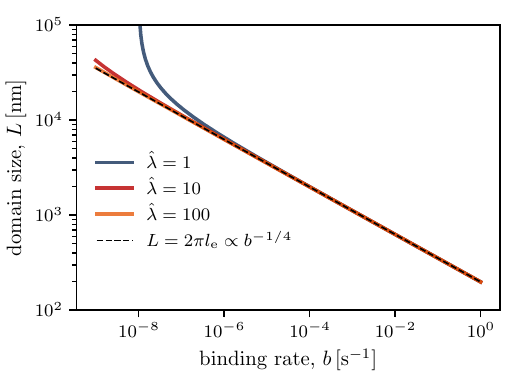}
    \caption{Domain size $L$ versus the binding rate $b$ for three different values of the dynamical asymmetry $\hhl$. The black dashed line gives the exchange-limited domain size. 
}
    \label{fig:real_unit}
\end{figure}

Figure~\ref{fig:real_unit} shows the steady-state
domain size $L$ as a function of the binding rate $b$ for three values of
the dynamic asymmetry $\hhl$. In the exchange-limited regime (high
$\hhl$), binding-release kinetics set the scale and $L \propto b^{-1/4}$.
Because $b$ is governed by motor abundance and cargo
affinity~\cite{gilliesCargoAdaptorIdentity2025,hummelSpecificKIF1AadaptorInteractions2021},
a hundredfold change in $b$ shifts $L$ by only $100^{1/4}\approx 3.2$, yet
sweeping $b$ across its physiological range moves $L$ continuously over a broad range up to the micron scale. This sublinear sensitivity makes the selected
size robust to fluctuations in motor expression while keeping it broadly
tunable.

The dynamic asymmetry determines where in parameter space this control operates. Because
$e_c = \hhs/\hhl^2$, larger dynamic asymmetry $\hhl$ lowers the critical exchange ratio and
shifts the transition to macrophase separation to smaller binding rates. As $\hhl$ decreases,
for instance when motor activity is
reduced~\cite{firestoneSmallmoleculeInhibitorsAAA2012,kumarMotorMutantsBring2014},
the system leaves the exchange-limited regime and condensates grow larger,
recovering bulk phase separation once transport becomes negligible.
Size control thus rests on two distinct handles: The binding rate $b$ sets the domain
size within a given transport regime, while the dynamic asymmetry $\hhl$ fixes the location of
the critical point $e_c$ and hence the
threshold binding rate $b_c = e_c\cdot s$ above which microphase separation occurs.

The resulting sizes span the range of biomolecular condensates observed in
cells, from neuronal RNA transport granules of a few hundred
nanometers~\cite{batishNeuronalMRNAsTravel2012} to micron-scale bodies such as
nucleoli and germline P granules~\cite{Kiebler:NatRevNeuroScience:2024}.
The largest values in Fig.~\ref{fig:real_unit}, of order hundreds of microns, reflect
the critical divergence as $e\to e_c$: on approach to the critical point $L$ grows
as $(e-e_c)^{-1/2}$, which is why the $\hhl=1$ branch turns upward at small $b$. Micron-scale domains, by contrast, lie well inside the exchange-limited regime and follow the $b^{-1/4}$ law. The mechanism thus enables both small and large condensates within a single framework. 

It should be noted that Figure~\ref{fig:real_unit}
must be interpreted as an upper bound for domain sizes if active transport takes place, the actual condensate sizes can be lower. Domain sizes of tens of nanometers, as calculated for the lowest binding rates, are not possible in most cells. In this regime, the actual sizes will be constrained by additional factors such as geometry or the amount of available material.

\begin{table}[ht]
\centering
\small
\begin{tabular}{lll}
\hline\hline
Para- & Definition & Description \\
meter & & \\
\hline
$\hhs$   & $s\,g\rO/(M k_B T)$ &(scaled release rate) \\
$\hhl$        & $\lambda \rO/(M k_B T)$ &(dynamic asymmetry) \\
$e$        & $b/s$ &(binding–release ratio) \\
$e_c$      &  $\hhs/\hhl^2$
&(critical $e$ for microphase onset) \\
$\ldinf$&  $\displaystyle (Mg k_BT/b \rO)^{1/4}$ 
&(exchange-limited domain size) \\ 
$L$ & $2\pi\,\ldinf (1- \sqrt{e_c/e})^{-1/2}$ \, &(general domain size) \\
\hline\hline
\end{tabular}
\caption{Summary of key parameters and results.}
\label{tab:parameters}
\end{table}
\section{Discussion}
We have demonstrated that unbiased motor-mediated protein transport along cytoskeletal filaments acts as an effective long-range repulsion
between proteins, shifting the LLPS spinodal and arresting coarsening
at a tunable length scale. The physical origin of this repulsion is
the nonlocal redistribution of motor-bound proteins: whenever a
condensate grows beyond the transport length $\Delta l_\text{a} =
\sqrt{\lambda/s}$, bound proteins are carried away faster than
diffusion can replenish them, creating a flux that opposes further
growth. This mechanism is captured by the simple scaling law $L
\propto b^{-1/4}$ (Table~\ref{tab:parameters}), which holds across a
wide range of exchange ratios and is robust to the presence of
noise.

The defining feature of this mechanism is that it leaves the proteins
untouched. Reaction-diffusion models of coarsening
arrest~\cite{weberPhysicsActiveEmulsions2019} control size by
chemically converting phase-separating molecules into a soluble,
non-condensing species; here, a motor-bound protein still retains its chemical identity, fully capable of condensation once released. Size control
therefore requires no engineered interaction domains or enzymatic
modifications, and can in principle act on any condensate whose
components can be recruited to a motor. The threshold for this regime
is strikingly low. Because the critical exchange ratio $e_c =
\hhs/\hhl^2$ scales as the inverse square of the dynamic asymmetry
between transport and free diffusion, and motors move cargo orders of
magnitude faster than proteins diffuse, only a minute motor-bound
fraction is needed to arrest coarsening. The accompanying sublinear
sensitivity (a hundredfold change in binding rate shifts $L$ by just
a factor of $\sim 3$) makes the resulting size robust to
fluctuations in motor expression and ATP availability.

Transport also controls shape, not just size. In a polarized
cytoskeleton, anisotropic redistribution elongates spherical
condensates into cylinders and lamellae once the anisotropy exceeds a modest threshold, with no change to the underlying phase diagram. This
separation of geometric from thermodynamic control is, to our
knowledge, distinctive: a cell could in principle reprogram condensate
morphology purely by remodeling its filament network. The prediction is
directly testable, since recruiting phase-separating proteins to motors
or tuning motor
activity~\cite{cochardCondensateFunctionalizationMicrotubule2023,gengKinesin3KIF1CUndergoes2024,firestoneSmallmoleculeInhibitorsAAA2012,kumarMotorMutantsBring2014}
should shift both size and shape in the predicted directions, and
polarized cell types such as neurons offer a natural setting in which
to look for transport-shaped condensates. The same principles suggest a
design strategy for synthetic active emulsions \cite{daiEngineeringSyntheticBiomolecular2023}, where pairing a
phase-separating scaffold with a tunable motor module would yield
condensates whose architecture is programmable independently of the thermodynamics.

Our work establishes a potential mechanism of size selection in
biomolecular condensates: unbiased motor-driven transport is by itself
sufficient to arrest coarsening and select a finite condensate size, from a few hundred nanometres up to the micron scale. Whether some cells actually exploit it is an open question, but given the simplicity of the mechanism it seems not unlikely, and the predictions above, the size law $L\propto b^{-1/4}$, the onset threshold $e_c$, and the anisotropy-driven shape transitions can be tested {\em in vitro}.
The mechanism is particularly suited to stabilize larger domains. If active, it sets an upper limit to the size of condensates, and it may act in concert with other local geometric factors or size control mechanisms that provide further constraints.

Our model isolates the transport mechanism in its simplest form, treating
redistribution as effective diffusion along a disordered network. This makes the
scaling laws transparent, and it also identifies the ingredients whose interplay
with transport would be interesting to explore next. Directed transport,
filament architecture~\cite{lombardoMyosinVaTransport2019} and motor
processivity~\cite{casonSelectiveMotorActivation2022} all shape how cargo
redistributes in structured cytoskeletons, and should modify the selected size
and morphology in ways the present framework can accommodate with some adjustments, such as adding convective flow terms. Collective active
flows from cytoskeletal contraction and microtubule sliding generate large-scale
advection in the
cytoplasm~\cite{lapradeCoarseningBiomimeticCondensates2026} that may reinforce
or compete with the redistribution described here, and hydrodynamic coupling
between
condensates~\cite{doanDiffusiophoresisPromotesPhase2024,berryPhysicalPrinciplesIntracellular2018}
becomes relevant at high volume fractions. The scaling laws derived here provide
a quantitative baseline against which such effects can be measured, and a
starting point for studying specific cellular implementations of
transport-controlled condensate organization.

\begin{acknowledgments}
This research was funded by the Deutsche Forschungsgemeinschaft (DFG, Germany) -- Project numbers 465145163 (CRC 1552); 248882694; 464588647 (CRC 1551). L.S.S.\ acknowledges support by M$^{\mathrm{3}}$ODEL and the Forschungsinitiative des Landes Rheinland-Pfalz. The authors gratefully acknowledge the computing time granted on the MOGON 2 supercomputer at Johannes Gutenberg University Mainz (https: //hpc.uni-mainz.de).
\end{acknowledgments}
\noindent\textit{Data availability}---The data that support the findings of this article are openly available in the zenodo repository at \url{https://doi.org/10.5281/zenodo.21780549}

\noindent\textit{Author declaration:} The authors declare no competing interests.

\appendix
\section{Model and Simulation Methods}
\label{sec:appendix}
The dimensionless dynamical equations governing the free and transported protein
densities, $\phi_1$ and $\phi_2$, are given by
\begin{align}
\partial_{\hht} \phi_1 &= \hat{\nabla}^2 
  \frac{\delta \hhF[\phi]}{\delta \phi_1} 
   - e\,  \hhs\phi_1 + \hhs \phi_2
   + \hat{\eta}_e, \label{eq:phi1}\\
\partial_{\hht} \phi_2 &=  \hat{\nabla} \bhhl\:  \hat{\nabla} \phi_2 
   + e\,\hhs\phi_1 - \hhs\phi_2
   - \hat{\eta}_e \label{eq:phi2}
\end{align}
where $\phi = \phi_1 + \alpha\phi_2$ is the total order parameter and
$\alpha = 0$ for the simplified model (used in the first part of the linear stability analysis) or
$\alpha = 1$ for the full model. All variables are
dimensionless; hatted symbols denote rescaled quantities. The parameter $e = b/s$
is the binding-release ratio, $\hhs$ is the scaled release rate, and $\bhhl$ is
the rescaled transport tensor. In the isotropic limit, where the
cytoskeleton forms a disordered meshwork with no preferred orientation,
$\bhhl = \hhl\,\boldsymbol{I}$ (with $\boldsymbol{I}$ the identity
tensor); anisotropic transport arising from aligned filament architectures is
modeled by a diagonal tensor, $\bhhl = \mathrm{diag}(\hhl_\parallel,
\hhl_\perp, \hhl_\perp)$ with $\hhl_\parallel > \hhl_\perp$. The dimensionless
free energy functional reads
\begin{equation}
    \hhF [\phi]=    
     \int_{\hat{V}} \text{d}^3\hhr\,
    \Bigl[
      \frac{\tau_2}{2}(\phi-\phi_c)^2
      +\frac{\tau_4}{4}(\phi-\phi_c)^4
      +\frac{1}{2}|\hat{\nabla}\phi|^2
    \Bigr]
\label{eq:F_with_phi2}
\end{equation}
where $\tau_2$ and $\tau_4$ control the bulk thermodynamics,
$\phi_c := \rho_c/\rO - 1$ sets the critical composition, and the
gradient term penalizes interfaces. 
The rescaled noise is given by
\begin{multline}
\hat{\eta}_e(\hat{\mathbf{r}}, t) = C_0\sqrt{\hat{s}}
  \left(\sqrt{e\phi_1 + \frac{e}{1+e}}\;\hat{\zeta}_1(\hat{\mathbf{r}},t)
  \right. \\
  \left. - \sqrt{\phi_2 + \frac{e}{1+e}}\;\hat{\zeta}_2(\hat{\mathbf{r}},t)
  \right)
\label{eq:noise_rescaled}
\end{multline}
with independent Gaussian white-noise fields satisfying
$\langle \hat{\zeta}_i (\hbr,\hht)\rangle \equiv 0$ and
$\langle \hat{\zeta}_i (\hbr,\hht) \hat{\zeta}_j (\hbr',\hht') \rangle
= \delta_{ij} \, \delta(\hbr-\hbr') \, \delta(\hht-\hht')$. The noise amplitude
$C_0 = g^{-3/4}/\sqrt{\rO}$ sets the strength of the fluctuations; here we
explore $C_0\in\{0, 0.1, 0.2\}$ to assess noise effects.

We solve Eqs.~(\ref{eq:phi1})–(\ref{eq:phi2}) using a semi-implicit
pseudo-spectral (FFT-based) method with periodic boundary conditions. All stiff
linear operators (e.g., $\hat{\nabla}^4\phi$, $\hat{\nabla}^2\phi$, $\phi$) are
treated implicitly in Fourier space to ensure numerical stability, while
nonlinear terms (e.g., $\hat{\nabla}^2[(\phi - \phi_c)^3]$) and noise
contributions are advanced explicitly. The stochastic terms follow Itô
discretization, consistent with Eq.~\ref{eq:noise_rescaled}. Full numerical details
are provided in the SI. 

Unless stated otherwise, we use a cubic computational domain of side
$128\,\lO$ with grid spacing $h=\lO$, yielding $128^3$ lattice points. Time is
discretized with fixed step $\Delta\hht = 0.5$. Initial conditions are spatially
homogeneous with small random perturbations:
$\phi_i(\hbr,0)\in[-10^{-3},10^{-3}]$ drawn from a uniform distribution, with
mean $\langle\phi_i\rangle=0$. We fix $\phi_c = 0.4$ and $\tau_4=0.4$, and vary
$\tau_2$ to control the quench depth $\Delta\tau := \tau -
\tau_{c1}^{\rm F}(e)$ (where $\tau = \tau_2 + 3\tau_4\phi_c^2$), measured
against the spinodal of the full model; for the simplified-model runs of SI~Sec.~V it is measured against $\tau_{c1}(e)$ instead, so that
$\Delta\tau=0$ always denotes the marginal state of the model being simulated.
For the isotropic transport simulations (Fig.~\ref{fig:3}), we set $\hhs=1$ and
$\hhl=10$, corresponding to a critical exchange ratio $e_c=0.0102$ for the
full model ($e_c=\hhs/\hhl^2=0.01$ for the simplified one). The exchange ratio
$e$ is varied from $0.02$ to $0.1$ to span the macro-to-microphase transition;
noise amplitude $C_0$ is varied in $\{0, 0.1, 0.2\}$ to probe stochastic effects.
Simulations are run up to $\hht=5\times10^4$ to ensure steady-state domain
statistics or confirm coarsening behavior.

To quantify the characteristic length scale, we compute the structure factor
$S(\mathbf{q}) = \langle|\hat{\phi}_1(\mathbf{q})|^2\rangle$ from the Fourier
transform of the free-protein density field and perform angular averaging to
obtain $S(q) = \langle S(\mathbf{q})\rangle_{|\mathbf{q}|=q}$. The dominant
length scale $L$ is extracted by fitting $S(q)$ to the empirical peak form
\[
S(q) = S_{\max}\exp\Bigl\{1 - \bigl[1 + \bigl((q- \qO)/\sigma\bigr)^2\bigr]^{1-\beta}\Bigr\},
\]
introduced in Ref.~\cite{qiaoEmpiricalMethodCharacterize2022}, which captures
both the central peak at wavenumber $ \qO$ and the high-$q$ power-law or
exponential decay. The domain size is then $L = 2\pi/ \qO$. Fits are performed
over a wavenumber window that includes the peak and at least one oscillation;
uncertainty estimates reflect the covariance of the fit parameters.
Time-dependent domain-size evolution (SI, Fig.~S3) confirms
$L(t)\propto \hht^{1/3}$ for macrophase coarsening ($e < e_c$) and plateau
behavior for microphase arrest ($e > e_c$).

\bibliographystyle{apsrev4-2}
\bibliography{Refs}

\end{document}


\newcommand{\hhr}{\hat{r}}
\newcommand{\hht}{\hat{t}}
\newcommand{\hbr}{\hat{\mathbf{r}}}
\newcommand{\hhl}{\hat{\lambda}}
\newcommand{\hhs}{\hat{s}}
\newcommand{\hhF}{\hat{\mathcal F}}
\newcommand{\hhFlin}{\hat{\mathcal F}^{\text{lin}}}
\newcommand{\tphi}{\tilde{\boldsymbol{\phi}}}
\newcommand{\rO}{\rho_{_0}}
\newcommand{\rone}{\rho_{_1}}
\newcommand{\rtwo}{\rho_{_2}}
\newcommand{\hhq}{\hat{q}}
\newcommand{\hbq}{\hat{\mathbf{q}}}
\newcommand{\hhFeff}{\hat{\mathcal F}^{\text{eff}}_t}
\newcommand{\eg}{\textit{e.g.}}
\newcommand{\lO}{\ell_{_0}}
\newcommand{\tO}{t_{_0}}
\newcommand{\ldinf}{l_\text{e}}
\newcommand{\Tr}{\text{Tr}}
\newcommand{\bhhl}{\hat{\boldsymbol{\lambda}}}
\title{Supplementary Material: Active Transport as a Mechanism of Microphase Selection in Biomolecular Condensates}
\author{Le Qiao}
\email{le.qiao@uni-mainz.de}
\affiliation{Institute of Physics, Johannes Gutenberg University Mainz, D55099 Mainz, Germany}
\author{Peter Gispert}
\affiliation{Institute of Physics, Johannes Gutenberg University Mainz, D55099 Mainz, Germany}
\author{Lukas S. Stelzl}
\affiliation{Institute of Molecular Physiology, Johannes Gutenberg University Mainz, D55099 Mainz, Germany}
\affiliation{Institute of Molecular Biology (IMB), Mainz, Germany}
\author{Friederike Schmid}
 \email{friederike.schmid@uni-mainz.de}
\affiliation{Institute of Physics, Johannes Gutenberg University Mainz, D55099 Mainz, Germany}
\maketitle
\section{Theoretical Model}

\subsection{Rescaled Dimensionless Equations}
We write the total protein concentration at position $\mathbf r$ as
\begin{equation}
      \rho(\mathbf r,t)
  = \rone(\mathbf r,t) + \rtwo(\mathbf r,t),
  \qquad
  \rO =  \bar\rho = \text{const.}
\end{equation}
In the homogeneous steady state, all gradient terms vanish and the
dynamics reduce to the reaction part, so the stationary condition is
$b\,\bar\rho_1 = s\,\bar\rho_2$ (from $\partial_t\rone=-b\rone+s\rtwo=0$
in the uniform case). Together with the conserved total density
$\bar\rho_1+\bar\rho_2=\rO$, this fixes the concentrations of the two species uniquely,
\begin{equation}
      \bar \rho_1 = \rO\,\frac{s}{b+s},
  \quad
  \bar\rho_2 = \rO\,\frac{b}{b+s}.
\end{equation}
The coupled dynamics in the $\rho$–variables are given by
\begin{align}
  \partial_t  \rone
  &= M\,\nabla^2\!\Bigl(\frac{\delta \mathcal F}{\delta \rone}\Bigr)
   -b\, \rone + s\,\rtwo
   + \eta_e,
  \label{eq:rho1}\\
  \partial_t \rtwo
  &= \lambda\,\nabla^2\rtwo
   +b\,\ \rone - s\,\rtwo
   - \eta_e.
  \label{eq:rho2}
\end{align}
with the active exchange noise
\begin{equation}
      \eta_e(\mathbf r,t)
  = \sqrt{b\, \rone}\,\zeta_1(\mathbf{r},t)
    - \sqrt{s\,\rtwo}\,\zeta_2(\mathbf{r},t)
\end{equation}
where the $\zeta_i(\mathbf{r},t)$ are Gaussian distributed white noise fields that satisfy
\begin{equation}
\label{eq:noise_fdr}
  \langle \zeta_i(\mathbf r,t) \rangle = 0, \qquad
  \langle \zeta_i(\mathbf r,t)\,\zeta_j(\mathbf r',t')\rangle
  = \delta_{ij}\,\delta(\mathbf r-\mathbf r')\,\delta(t-t').
\end{equation}
This exchange noise is the chemical-Langevin (shot) noise of the
binding/release reactions, constructed following Gillespie's chemical
Langevin equation~\cite{gillespieChemicalLangevinEquation2000}, and is
interpreted in the It\^o sense~\cite[Sec.~2.3, p.~40]{paulbaschnagel_book}. It accounts for the stochastic character of binding/release events and becomes important if the concentrations of bound or free species are small, such that a treatment in terms of deterministically evolving continuous fields is no longer appropriate. The precise form was derived by Gillespie for the case of bulk reactions\cite{gillespieChemicalLangevinEquation2000}, based on the assumption that individual reactions are independent from each other. When applied to spatially varying concentration fields, the
same assumption leads to the spatial delta correlation, $\delta(\mathbf r-\mathbf r')$, in Eq. (\ref{eq:noise_fdr}). We will see below that the amplitude of the noise scales with $1/\sqrt{\rO}$ in appropriately reduced units, such that the deterministic limit is reached at $\rO \to \infty$.
 
Equations~\eqref{eq:rho1}--\eqref{eq:rho2} do not include conserved noise: we retain only the athermal exchange noise, so that $\rO \to \infty$ defines
the deterministic baseline against which the stochastic contribution is measured.
This distinction matters for Sec.~\ref{sec:noise}, where the fact that
$\eta_e$ enters outside the Laplacian is precisely what produces the
$\hhq^{-4}$ tail, in contrast to the $\hhq^{-2}$ that conserved noise
would give.

We assume bound proteins ($\rtwo$) to participate in interactions in the same way as free particles ($\rone$), 
just with a different dynamics, and assume a simple Ginzburg-Landau type interaction potential
\begin{equation}
     {\mathcal F}[\rho]
\rem{  =\rO \frac{F}{k_BT} = \rO^2}
   = \rO \: k_B T
    \int_V \text{d}^3r\,
    \Bigl[
      \tfrac{\tau_2}{2 \rO^2}(\rho-\rho_c)^2
      +\tfrac{\tau_4}{4 \rO^4}(\rho-\rho_c)^4
      +\tfrac{g}{2 \rO^2}|\nabla\rho|^2
    \Bigr],
\end{equation}

\medskip
\noindent
To derive the rescaled equations, we define normalized order parameter
\begin{equation}
\phi_i(\mathbf r,t) = \frac{\rho_i(\mathbf r,t)-\bar\rho_i}{\rO}, \qquad \bar\phi_i = 0 .
\end{equation}

Since $\partial_t\phi_i = \frac1{\rO}\,\partial_t\rho_i$, replacing every $\rho_i$
in \eqref{eq:rho1}–\eqref{eq:rho2} yields
\begin{align}
  \partial_t\phi_1
  &= \frac{M}{\rO^2}\,\nabla^2 \frac{\delta \mathcal F}{\delta\phi_1}
     -b\,\phi_1 + s\,\phi_2
     + \frac{\eta_e}{\rO},
  \label{eq:phi1orig}\\
  \partial_t\phi_2
  &= \lambda\,\nabla^2 \phi_2
     +b\,\phi_1 - s\,\phi_2
     - \frac{\eta_e}{\rO}.
  \label{eq:phi2orig}
\end{align}
The noise rescales as
\begin{equation}
      \frac{\eta_e (\mathbf{r},t)}{\rO}
  = \frac{1}{\sqrt{\rO}}\Bigl[
     \sqrt{b\,\phi_1 + \tfrac{b\,s}{b+s}}\;\zeta_1(\mathbf{r},t)
     -\sqrt{s\,\phi_2 + \tfrac{b\,s}{b+s}}\;\zeta_2(\mathbf{r},t)
   \Bigr].
   \label{eq:noise_exp}
\end{equation} 

We now rewrite these equations in terms of rescaled lengths
$\hhr = r/\ell_0$ and times $\hht = t/t_0$ with $\ell_0 = \sqrt{g}$ 
and $t_0 = \ (g \rO)/(M k_B T)$. Further defining the
dimensionless parameters (see main text)
$e = b/s$,  $\hhl = (\lambda \rO)/(M k_B T)$, 
$\hhs =(s g \rO)/(M \, k_B T)$, $C_0= g^{-3/4}/\sqrt{\rO}$,
and $\phi_c = \rho_c/\rO - 1$, 
the coupled dynamical equations then simplify to 
\begin{align}
\partial_{\hht} \phi_1 &= \hat{\nabla}^2 
  \frac{\delta \hhF[\phi_1+\phi_2]}{\delta \phi_1} 
   - e\,  \hhs\phi_1 + \hhs \phi_2
   + \hat{\eta}_e, \label{eq:phi1}\\
\partial_{\hht} \phi_2 &= \hhl \: \hat{\nabla}^2 \phi_2 
   + e\,\hhs\phi_1 - \hhs\phi_2
   - \hat{\eta}_e \label{eq:phi2}
\end{align}
In these equations, the rescaled free energy functional reads
\begin{equation}
    \hhF [\phi]=
  \frac{{\mathcal F}[\rO (1+\phi)]}{\rO \ell_0^3\: k_BT} =     
     \int_{\hat{V}} \text{d}^3\hhr\,
    \Bigl[
      \frac{\tau_2}{2}(\phi-\phi_c)^2
      +\frac{\tau_4}{4}(\phi-\phi_c)^4
      +\frac{1}{2}|\hat{\nabla}\phi|^2
    \Bigr]
\label{eq:F_with_phi2}
\end{equation}
and has the derivative
\begin{equation}
    \frac{\delta \hhF[\phi_1+\phi_2]}{\delta\phi_1}
  = \tau_2(\phi_1 + \phi_2-\phi_c) + \tau_4(\phi_1+\phi_2-\phi_c)^3 
  -  \hat{\nabla}^2(\phi_1+\phi_2).
  \label{eq:chem_pot}
\end{equation}

The rescaled noise in rescaled units, $\hat{\eta}_e$, is given by 
\begin{equation}
\hat{\eta}_e (\hbr,\hht)
 = C_0 \: \sqrt{\hhs} \: 
 \left( \sqrt{e \phi_1 + \frac{e}{1+e}} \; \hat{\zeta}_1 (\hbr,\hht)
 - \sqrt{\phi_2 + \frac{e}{1+e}} \; \hat{\zeta}_2(\hbr,\hht)
 \right)
\label{eq:noise_rescaled}
\end{equation}
with $\langle \hat{\zeta}_i (\hbr,\hht)\rangle \equiv 0$ and
$\langle \hat{\zeta}_i (\hbr,\hht) \hat{\zeta}_j (\hbr',\hht') \rangle 
= \delta_{ij} \, \delta(\hbr-\hbr') \, \delta(\hht-\hht')$. 

The above Eqs. \eqref{eq:phi1}-\eqref{eq:noise_rescaled} are used in our 3D simulations for the results shown in the main text. 

\section{Linear stability analysis}
\label{si:LSA}
\subsection{Effective free energy for the simplified model}
\label{sec:Feff}

The dynamics of Eqs.~\eqref{eq:phi1}--\eqref{eq:phi2} are not of relaxational
form, because the exchange terms are not derivable from a free energy. If the
transported density, $\rtwo$, is omitted from the free energy, however, they can be mapped onto an effective relaxational model, which is what connects the system to the Ohta--Kawasaki class and yields the closed-form spinodal quoted in the main text.
We construct that mapping here; the resulting functional is also the starting
point for the analysis of the noise in Sec.~\ref{sec:noise}.

Omitting $\rtwo$ from the free energy gives
\begin{equation}
  {\mathcal F}_1
  = \rO k_B T \int_V \mathrm d^3r\,
    \Bigl[
      \frac{\tau_2}{2\rO^2}(\rone-\rho_c)^2
      +\frac{\tau_4}{4\rO^4}(\rone-\rho_c)^4
      +\frac{g}{2\rO^2}|\nabla\rone|^2
    \Bigr],
\end{equation}
or in rescaled units
\begin{equation}
  \hhF_1[\phi_1]
  = \int_{\hat V}\mathrm d^3\hhr\,
    \Bigl[
      \frac{\tau_2}{2}(\phi_1-\phi_c)^2
      +\frac{\tau_4}{4}(\phi_1-\phi_c)^4
      +\frac{1}{2}|\hat\nabla\phi_1|^2
    \Bigr].
\end{equation}
Defining in addition
\begin{equation}
  \hhF_2[\phi_2] = \int_{\hat V}\mathrm d^3\hhr\,\frac{\hhl}{2e}\,\phi_2^2 ,
  \qquad
  \hhF_t[(\phi_1,\phi_2)] = \hhF_1[\phi_1] + \hhF_2[\phi_2],
\end{equation}
and rescaling the second field as
$\tphi=(\tilde\phi_1,\tilde\phi_2)=(\phi_1,\phi_2/\sqrt e)$ so as to symmetrize
the exchange terms, Eqs.~\eqref{eq:phi1}--\eqref{eq:phi2} take the matrix form
\begin{equation}
  \partial_{\hht}\tphi
  = \hat\nabla^2\frac{\delta\hhF_t}{\delta\tphi}
    + \mathbf A\,\tphi
    + \hat\eta_e{1\choose -1/\sqrt e},
  \qquad
  \mathbf A = \hhs\begin{pmatrix} -e & \sqrt e\\ \sqrt e & -1\end{pmatrix}.
  \label{eq:phi_combined}
\end{equation}
The factor $1/e$ in $\hhF_2$ is required by this rescaling: the functional
derivative in Eq.~\eqref{eq:phi_combined} is taken with respect to $\tphi$, and
$\hat\nabla^2\,\delta\hhF_2/\delta\tilde\phi_2 = \hhl\,\hat\nabla^2\tilde\phi_2$
reproduces the transport term of Eq.~\eqref{eq:phi2} only with this coefficient.
Thus $\hhF_2$ encodes the transport term rather than a thermodynamic free energy.
In these variables, one gets
\begin{equation}
  \hhF_t[\tphi] =
    \int_{\hat V}\mathrm d^3\hhr\,
    \Bigl[
      -\phi_c(\tau_2+\tau_4\phi_c^2)\,\tilde\phi_1
      +\frac{\tau_2+3\tau_4\phi_c^2}{2}\,\tilde\phi_1^2
      -\tau_4\phi_c\,\tilde\phi_1^3
      +\frac{\tau_4}{4}\tilde\phi_1^4
      +\frac{1}{2}|\hat\nabla\tilde\phi_1|^2
      +\frac{\hhl}{2}\,\tilde\phi_2^2
    \Bigr] + \mathrm{const}.
\end{equation}
The terms that are constant or linear in $\tilde{\phi}_1$ do not contribute to  the dynamical equations and can be omitted.
Using a Fourier representation for  the quadratic terms, the functional $\hhF_t[\tphi]$ can thus equivalently be    replaced by
 \begin{equation}
   \hhF_t'[\tphi] =
     \frac{1}{2{\hat{V}}} \sum_{\mathbf{q} \neq 0}
     \Bigl[
     (\tau_2 + 3 \tau_4 \phi_c^2 + q^2) \, |\tilde{\phi}_1(\mathbf{q})|^2
     + \hhl \: |\tilde{\phi}_2(\mathbf{q})|^2
     \Bigl]
     +
     \int_{\hat{V}} \text{d}^3\hhr\,
     \Bigl[
       - \tau_4 \phi_c \: \tilde{\phi}_1^3 
       + \frac{\tau_4}{4}\tilde{\phi}_1^4
     \Bigr].  
 \end{equation}
Here the Fourier transform is defined
via $\tilde{\phi}_i(\mathbf{q}) = \int_{\hat V}\mathrm d^3\hhr\, \tilde{\phi}_i(\hbr) \: \text{e}^{-i \mathbf{q}\cdot \hbr}$, which  implies
$\tilde{\phi}_i(\hbr) = \frac{1}{\hat V} \sum_{\mathbf{q}} \, \tilde{\phi}_i(\mathbf{q}) \: \text{e}^{i \mathbf{q}\cdot \hbr}$, and we have used Parseval's theorem with $\hat\nabla\to i\mathbf{q}$ producing the $q^2$
term.
 Next we exploit the identity

\begin{align}
  \mathbf A\,\tphi = \hat\nabla^2\frac{\delta\hhF_{\rm Coul}}{\delta\tphi}
  \qquad\text{for}\qquad
  \hhF_{\rm Coul}[\tphi]
  &= -\frac{1}{2}\iint \tphi^{\mathsf T}(\hbr')\,\mathbf A\,
     G(\hbr-\hbr')\,\tphi(\hbr)\,\mathrm d^3\hhr\,\mathrm d^3\hhr'\\
  &= -\frac{1}{2\hat V}\sum_{\hbq\neq0}
     \tphi^\dagger_{\hbq}\,\mathbf A\,G_{\hbq}\,\tphi_{\hbq},
\end{align}
with $G$ the Green's function of the Laplacian,
$\hat\nabla^2G(\hbr-\hbr')=-\delta(\hbr-\hbr')$, i.e.\ $G_{\hbq}=1/\hhq^2$. The
$1/\hhq^2$ kernel is the long-range Coulomb-like interaction generated by the
exchange. Equation~\eqref{eq:phi_combined} therefore becomes
\begin{equation}
  \partial_{\hht}\tphi
  = \hat\nabla^2\frac{\delta\hhF_{\rm eff}}{\delta\tphi}
    + \hat\eta_e{1\choose -1/\sqrt e},
  \label{eq:phi_combined_2}
\end{equation}
with the virtual free energy $\hhF_{\rm eff}=\hhF_t'+\hhF_{\rm Coul}$,
\begin{align}
  \hhF_{\rm eff} &=
    \frac{1}{2\hat V}\sum_{\hbq\neq0}
      \tphi^\dagger_{\hbq}\,\mathbf B_{\hbq}\,\tphi_{\hbq}
    + \int_{\hat V}\mathrm d^3\hhr\,
      \Bigl[-\tau_4\phi_c\,\tilde\phi_1^3+\frac{\tau_4}{4}\tilde\phi_1^4\Bigr],\\
  \text{where}\quad
  \mathbf B_{\hbq} &= \begin{pmatrix}
    \tau_2+3\tau_4\phi_c^2+\hhq^2+e\hhs/\hhq^2 & -\sqrt e\,\hhs/\hhq^2\\
    -\sqrt e\,\hhs/\hhq^2 & \hhl+\hhs/\hhq^2
  \end{pmatrix}.
  \label{eq:bmatrix}
\end{align}
We stress that Eq.~\eqref{eq:phi_combined_2} is not an equilibrium model with
free energy $\hhF_{\rm eff}$: the noise violates the fluctuation--dissipation
relation, so minimizing $\hhF_{\rm eff}$ need not give the correct steady state.

\subsection{Spinodal of both models}
\label{sec:spinodal}

Neglecting the noise, the homogeneous solution of Eq.~\eqref{eq:phi_combined_2} is $\tphi\equiv0$. We now determine when it first
becomes unstable, treating the simplified and the full model together. To do so
we let the free energy depend on
\begin{equation}
  \phi := \phi_1 + \alpha\,\phi_2,
  \qquad
  \alpha =
  \begin{cases}
    0 & \text{simplified model, }\hhF[\phi_1],\\
    1 & \text{full model, }\hhF[\phi_1+\phi_2],
  \end{cases}
\end{equation}
and carry $\alpha$ through the calculation. Linearizing the chemical potential of
Eq.~\eqref{eq:chem_pot} about the homogeneous state,
\begin{equation}
  \frac{\delta\hhF}{\delta\phi_1}\;\approx\;(\tau-\hat\nabla^2)\,\phi,
  \qquad \tau := \tau_2+3\tau_4\phi_c^2 ,
\end{equation}
and Fourier transforming, Eqs.~\eqref{eq:phi1}--\eqref{eq:phi2} become
$\partial_{\hht}\bm\phi_{\hbq}=-\mathbf M_{\hbq}\bm\phi_{\hbq}$ with
$\bm\phi_{\hbq}=(\phi_1,\phi_2)^{\mathsf T}$ and
\begin{equation}
  \mathbf M_{\hbq}
  = \hhq^2\begin{pmatrix}
    (\tau+\hhq^2)+e\hhs/\hhq^2 & \alpha(\tau+\hhq^2)-\hhs/\hhq^2\\[1mm]
    -e\hhs/\hhq^2 & \hhl+\hhs/\hhq^2
  \end{pmatrix}.
  \label{eq:si_M}
\end{equation}
For $\alpha=0$ the rescaling of Sec.~\ref{sec:Feff} renders
$\mathbf M_{\hbq}/\hhq^2$ symmetric and one recovers $\mathbf B_{\hbq}$ of
Eq.~\eqref{eq:bmatrix}. For $\alpha=1$, no rescaling achieves this, because $M_{12}$
carries the free-energy term $\hhq^2(\tau+\hhq^2)$ while $M_{21}$ does not:
the transported density enters $\hhF$ and hence drives $\phi_1$, but $\hhF$ does
not drive $\phi_2$, which moves by motor transport rather than driven by a
chemical-potential gradient. No effective free energy exists in that case, providing a precise sense in which the full model lies further from equilibrium.

The stability criterion itself requires no symmetry. For a $2\times2$ system the
eigenvalues satisfy $\mu_+\mu_-=\det\mathbf M_{\hbq}$ and
$\mu_++\mu_-=\operatorname{tr}\mathbf M_{\hbq}$, and both have positive real part
if and only if $\det\mathbf M_{\hbq}>0$ and $\operatorname{tr}\mathbf M_{\hbq}>0$.
Evaluating the trace on the spinodal, where
$\tau+\hhq^2=-e\hhs\hhl/[\hhl\hhq^2+\hhs(1+\alpha e)]$, we find
\begin{equation}
  \operatorname{tr}\mathbf M_{\hbq}
  = \hhl\hhq^2+\hhs(1+\alpha e)
    -\frac{e\hhs\hhl\hhq^2}{\hhl\hhq^2+\hhs(1+\alpha e)}\;>\;0
\end{equation}
for the parameters used. Thus no oscillatory (Hopf) instability occurs and the
marginal-stability condition is $\det\mathbf M_{\hbq}=0$ in both models. For
$\alpha=0$, this is equivalent to $\det\mathbf B_{\hbq}=0$, since
$\det\mathbf B_{\hbq}=\det\mathbf M_{\hbq}/\hhq^4$; explicitly, the $\hhq^{-4}$
terms cancel and
\begin{equation}
  \det\mathbf B_{\hbq}
  = \hhl\hhq^2 + (\tau\hhl+\hhs) + \frac{\hhs(\tau+e\hhl)}{\hhq^2},
  \label{eq:detB}
\end{equation}
a form we shall use in Sec.~\ref{sec:noise}.

Writing $X:=\hhq^2(\tau+\hhq^2)$, the determinant is
\begin{equation}
  \det\mathbf M_{\hbq}
  = X\bigl[\hhl\hhq^2+\hhs(1+\alpha e)\bigr] + e\,\hhs\,\hhl\,\hhq^2 .
  \label{eq:si_det}
\end{equation}
The entire $\alpha$ dependence thus enters through the single substitution
$\hhs\to\hhs(1+\alpha e)$. Solving $\det\mathbf M_{\hbq}=0$ for $\tau$ gives
$\tau(\hhq)=-\hhq^2-e\hhs\hhl/[\hhl\hhq^2+\hhs(1+\alpha e)]$; its $\hhq\to0$
limit is the macrophase branch, and its maximum over $\hhq$ is the microphase
branch together with the wave number selected there,
\begin{align}
  \tau_{c0}(e) &= -\frac{e\hhl}{1+\alpha e}, \label{eq:si_tauc0}\\
  \tau_{c1}(e) &= -2\sqrt{e\hhs}+\frac{\hhs(1+\alpha e)}{\hhl},
    \label{eq:si_tauc1}\\
  \hhq_0^2(e)  &= \sqrt{e\hhs}-\frac{\hhs(1+\alpha e)}{\hhl}.
    \label{eq:si_q0}
\end{align}
If $\alpha < \hhl^2/4 \hhs$, the two branches meet at a transition point $e_c$, defined via the implicit equation 
\begin{equation}
1 + \alpha \, e_c = \sqrt{e_c} \: \hhl/\sqrt{\hhs},
\end{equation}
where $\hhq_0\to0$ and the
characteristic domain size diverges. For $e$ below that point, only macrophase separation is possible ($ \hhq_0^2(e)$ becomes negative). Above it, the finite-$\hhq$ (microphase) mode is crossed first as $\tau$ is lowered, and the system microphase separates.

\subsection{Comparison of the two models}
\label{sec:compare}
For the simplified model ($\alpha=0$), Eqs.~\eqref{eq:si_tauc0}--\eqref{eq:si_q0}
reduce to $\tau_{c0}=-e\hhl$ and $\tau_{c1}=\hhs/\hhl-2\sqrt{e\hhs}$, meeting at
$e_c=\hhs/\hhl^2$, $\tau_c=-\hhs/\hhl$. In unrescaled units
Eq.~\eqref{eq:si_q0} reads
$q_0^2=\ell_0^{-2}\bigl(\sqrt{e\hhs}-\hhs/\hhl\bigr)$, the characteristic
wave number quoted in the main text.

For the full model ($\alpha=1$), microphase separation is only possible if $\hhs/\hhl^2<1/4$. The transition point between macro- and microphase separation shifts to higher values of $e_c$ according to
\begin{equation}
    \frac{\delta \sqrt{e_c}}{\sqrt{e_c}} :=
    \frac{\sqrt{e_c}|_{\alpha = 1} - \sqrt{e_c}|_{\alpha = o}}
    {\sqrt{e_c}|_{\alpha = 1}} = \frac{\sqrt{\hhs}}{\hhl}.
\end{equation}
The spinodals are shifted upwards by
\begin{equation}
   \Delta_0 := \tau_{c0}\big|_{\alpha=1}-\tau_{c0}\big|_{\alpha=0}
  = \frac{e^2 \hhl}{1+e}\;>\;0  \quad \text{and} \qquad
  \Delta_1 := \tau_{c1}\big|_{\alpha=1}-\tau_{c1}\big|_{\alpha=0}
  = \frac{\hhs\,e}{\hhl}\;>\;0 ,
  \label{eq:si_offset}
\end{equation}
so microphase separation sets in slightly earlier when $\phi_2$ is retained, since the
transported proteins then also contribute to the driving force for demixing. The
shift vanishes as $e\to0$ and grows linearly with $e$: for the parameters used,
$\Delta_1=2\times10^{-3}$ at $e=0.02$ and $5\times10^{-3}$ at $e=0.05$. The
selected $\hhq_0^2$ is smaller by the same amount, $\delta \hhq_0^2 = - \Delta_1$, and since $L=2\pi/\hhq_0$
implies $\delta L/L=-\tfrac12\,\delta(\hhq_0^2)/\hhq_0^2$, the relative domain size increases according to
\begin{equation}
  \frac{\delta L}{L}
  \simeq \frac{\hhs \, e}{2 \hhl \, \hhq_0^2\big|_{\alpha=0}}= \frac{\hhs\,e}{2\hhl\bigl(\sqrt{e\hhs}-\hhs/\hhl\bigr)} ,
  \label{eq:si_dL}
\end{equation}
which is about $2\%$ over the range studied in the simulations. The approximation underlying
$\hhF_{\rm eff}$ is therefore quantitatively controlled at the parameters considered in this work, and becomes least accurate on approaching the transition point $e = e_c$, where
$\hhq_0\to 0$ and the relative shift diverges.

\section{Numerical treatments}
On a cubic grid of spacing $\Delta \hat{h}=1$, time step $\Delta \hat{t}$,
and volume per cell $\Delta \hat{V}=1$, we discretize in Fourier space,
treating the stiff $k$-dependent linear terms implicitly and the remaining linear, non-linear and noise terms explicitly.
Let $\Phi_i^n(\mathbf k)$ be the Fourier transform of $\phi_i^n(\hat{\mathbf r})$.
Then with $\hat{\nabla}^2\to -k^2$, $\hat{\nabla}^4\to k^4$, the updates are
\begin{equation}
  \Phi_1^{n+1}
  = \frac{
      \Phi_1^n
    +\Delta \hat{t}\bigl[-e\hhs\,\Phi_1^n+
      \left(\hhs-k^2\,\tau_2-k^4\right)\,\Phi_2^n
      -k^2\,\tau_4 \,\text{FFT}\left[(\phi_1^n+\phi_2^n-\phi_c)^3\right]
      + \text{FFT}\left[\zeta_e^n\right]
    \bigr]
    }{
      1 + \Delta \hat{t}\bigl(k^4 + \,\tau_2\,k^2\bigr)
    },
\end{equation}
\begin{equation}
  \Phi_2^{n+1}
  = \frac{
      \Phi_2^n
    +\Delta \hat{t}\bigl[e\hhs\,\Phi_1^n -\hhs\,\Phi_2^n - \text{FFT}\left[\zeta_e^n\right]\bigr]
    }{
      1 + \Delta \hat{t}\bigl(\hhl\,k^2\bigr)
    }.
\end{equation}
Finally invert the FFT to obtain $\phi_i^{n+1}(\hat{\mathbf r})$ and repeat.

In our approximate model where $\phi_2$ is neglected in thermodynamics we have 
\begin{equation}
  \Phi_1^{n+1}
  = \frac{
      \Phi_1^n
    +\Delta \hat t\bigl[
      \hhs\,\Phi_2^n-e\hhs\,\Phi_1^n
      -k^2\,\tau_4\,\text{FFT}\left[(\phi_1^n-\phi_c)^3\right]
      +\text{FFT}\left[\zeta_e^n\right]
    \bigr]
    }{
      1 + \Delta \hat{t}\bigl(k^4 + \tau_2\,k^2\bigr)
    },
\end{equation}

\medskip\noindent{\bf Noise discretization in real space:}  On each grid cell of
volume $\Delta \hat V=1$ and time step $\Delta \hat t$, the unit white noises of
Eq.~\eqref{eq:noise_rescaled} are realized as
\begin{equation}
      \hat\zeta_i^n(\hat{\mathbf r})
  = \sqrt{\frac{1}{\Delta \hat{t}\,\Delta \hat{V}}}\;\Omega_i(\hat{\mathbf r},n), \quad i=1,2
\end{equation}
where $\Omega_i(\mathbf {\hat r},n)$ are independent standard Gaussians (zero
mean, unit variance) at each cell and time, the prefactor being the lattice
representation of $\delta(\hbr-\hbr')\,\delta(\hht-\hht')$.
The exchange noise entering the updates above is then assembled exactly as
in Eq.~\eqref{eq:noise_rescaled},
\begin{equation}
\zeta_e^n(\hbr)
  = C_0\sqrt{\hhs}\left(
      \sqrt{e\,\phi_1^n+\frac{e}{1+e}}\;\hat\zeta_1^n(\hbr)
    - \sqrt{\phi_2^n+\frac{e}{1+e}}\;\hat\zeta_2^n(\hbr)\right).
  \label{eq:zeta_e_discrete}
\end{equation}
The amplitude $C_0$ appears here only, through
Eq.~\eqref{eq:noise_rescaled}, and must not be included a second time in
$\hat\zeta_i^n$.

\section{Effect of the noise on the large-$\hhq$ tail of $S(\hhq)$}
\label{sec:noise}
The exchange noise leaves the position and width of the peak in $S(\hhq)$
essentially unchanged but lifts its large-$\hhq$ tail. We show here that this
follows from a linear analysis, and that the lift grows as $C_0^{2}$.

We first need the amplitude of the noise in the homogeneous state. From
Eq.~\ref{eq:noise_rescaled}, with $\hat\zeta_1,\hat\zeta_2$ independent unit
white noises, the cross term drops from
$\langle\hat\eta_e^{2}\rangle$ and the two variances add,
\begin{equation}
  \bigl\langle\hat\eta_e^{2}\bigr\rangle
  = C_0^{2}\hhs\left[\left(e\phi_1+\tfrac{e}{1+e}\right)
        + \left(\phi_2+\tfrac{e}{1+e}\right)\right]
  \;\xrightarrow[\;\phi_i\to0\;]{}\;
  \sigma_0^{2} = \frac{2\,C_0^{2}\,\hhs\,e}{1+e}\;\neq\;0 .
  \label{eq:si_sigma0}
\end{equation}
The mean exchange flux vanishes in the homogeneous state, since binding and
release balance, $b\bar\rho_1=s\bar\rho_2$; the fluctuations do not, because the
two reactions are independent stochastic events. Two independent event streams
of equal rate have zero mean difference but a variance equal to the sum of the
rates, which is the factor $2$ in Eq.~\eqref{eq:si_sigma0}.

At large $\hhq$ the off-diagonal elements of $\mathbf B_{\hbq}$
(Eq.~\ref{eq:bmatrix}) vanish as $\hhq^{-2}$, so the two fields decouple and
$\tilde\phi_1$ obeys a scalar Ornstein--Uhlenbeck equation,
\begin{equation}
  \partial_{\hht}\,\tilde\phi_1(\hbq)
  = -\,\Gamma(\hhq)\,\tilde\phi_1(\hbq) + \hat\eta_{\hbq}(\hht),
  \qquad
  \Gamma(\hhq) = \hhq^{2}\bigl(\tau + \hhq^{2}\bigr) \simeq \hhq^{4}.
\end{equation}
Its stationary variance is $\langle x^{2}\rangle=\sigma^{2}/2\Gamma$, so the
structure factor $S(\hhq)=\langle|\tilde\phi_1(\hbq)|^{2}\rangle$ is
\begin{equation}
  S(\hhq) \;\simeq\; \frac{\sigma_0^{2}}{2\,\Gamma(\hhq)}
  \;=\; \frac{C_0^{2}\,\hhs\,e}{(1+e)\,\hhq^{4}},
  \qquad \hhq^{2}\gg\hhs/\hhl \;\;\text{and}\;\; \hhq^{2}\gg|\tau| .
  \label{eq:si_tail}
\end{equation}

The interpretation is that the drive $\sigma_0^{2}$ is the same at every
wave number, because the noise is spatially white and enters outside the
Laplacian, whereas the damping grows with $\hhq^{4}$. Short-wavelength modes are
therefore forced as strongly as all the others but relax faster, settling at a
small residual level that decays only as a power law. The deterministic tail, by contrast, decays exponentially (the fitted exponent $1-\beta\to0.5$), such that the noise background dominates beyond a crossover wave number. Since $S_{\max}$ is set by the deterministic pattern and is essentially independent of $C_0$, the normalized tail rises in proportion to $\sigma_0^{2}\propto C_0^{2}$ while the peak is unchanged, as observed.

\section{Approximate Model and effect of noise}
\label{sec:appB}
\begin{figure*}[ht]
  \centering
\includegraphics[width=0.9\linewidth]{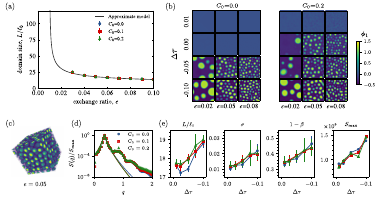}
\caption{Simulation results for the simplified two-field model (omitting
$\phi_2$ from the free energy functional), to be compared with the full model
(main text Sec.~V, Fig.~3). The quench depth is measured
against the spinodal of this model, $\Delta\tau:=\tau-\tau_{c1}(e)$, so that
$\Delta\tau=0$ lies on the spinodal.
(a)~Steady-state domain size $L/\lO$ versus exchange ratio $e$: theory (solid
line) and simulations at three noise strengths $C_0=0,\,0.1,\,0.2$ for
$\Delta\tau=-0.01$.
(b)~Midplane $\phi_1$ slices for $e=0.02,\,0.05,\,0.08$ without and with noise
($C_0=0$ and $C_0=0.2$) at time $\hht=5\times10^4$; rows correspond to four
quench depths $\Delta\tau=0.01,\,0.00,\,-0.05,\,-0.10$.
(c)~3D snapshot for $e=0.05$ at $\hht=5\times10^4$.
(d)~Angle-averaged structure factor $S(q)$ for $e=0.05$ at $\Delta\tau=-0.01$
and the three noise strengths, in semi-log representation; solid lines are fits
to the empirical form $S(q)\sim\exp\{1-[1+((q-q_0)/\sigma)^2]^{1-\beta}\}$.
(e)~Fitted domain size $L$, peak width $\sigma$, exponent $1-\beta$ and peak
amplitude $S_{\max}$ versus quench depth $\Delta\tau$ for $e=0.05$.
Parameters are $\hhs=1$, $\hhl=10$, $\phi_c=0.4$, $\tau_4=0.4$.}

  \label{fig:A1}
\end{figure*}

\noindent\textbf{Comparison with the full model:~} Figure~\ref{fig:A1} shows
simulation results for the simplified two-field model (omitting $\phi_2$ from
the free energy) used in the linear stability analysis. Here the quench depth is
measured against the spinodal of each model, so that $\Delta\tau=0$ corresponds
in both cases to the respective marginal state. Compared in this way, the two
models agree closely: the predicted domain sizes are indistinguishable on the
scale of panel~(a), and at $\Delta\tau=0$ without noise neither model develops
structure within the simulated time, as expected since the critical mode is
marginal in both.

The residual difference is the shift of the spinodal derived in
Sec.~\ref{si:LSA} C, $\delta\tau_{c1}=\hhs e/\hhl$, which also shifts the selected wave number and increases the domain sizes in the full model by
$\delta L/L\simeq \hhs e/\bigl[2\hhl\bigl(\sqrt{e\hhs}-\hhs/\hhl\bigr)\bigr]$.
This amounts to about $2\%$ over the range studied, consistent with the fitted domain sizes at $\Delta\tau=-0.1$, $L/\lO\simeq18.9$ in panel~(e) versus $19.3$ in Fig.~3(e) in the main text. The approximation underlying $\hhFeff$ is
therefore quantitatively controlled at the quench depths used here, becoming
least accurate on approach to the transition point to macrophase separation, where $q_0\to0$ and the
relative shift diverges.

\begin{figure}[t]
\includegraphics[width=0.6\linewidth]{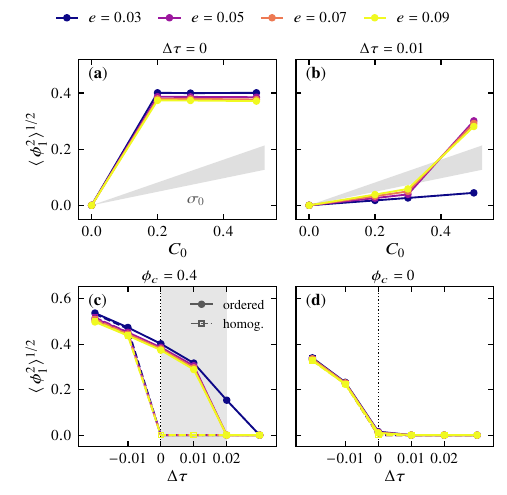}
\caption{Effect of the exchange noise at and above the spinodal for the
simplified model ($\phi_2$ omitted from the free energy). Colour encodes the
exchange ratio $e$ in all panels.
(a),(b)~RMS amplitude $\langle\phi_1^2\rangle^{1/2}$ at $\hht=5\times10^4$ as a
function of the noise strength $C_0$, for quench depths $\Delta\tau=0$ (on the
spinodal) and $\Delta\tau=0.01$ (above it). All runs start from a homogeneous
state with a random perturbation of amplitude $10^{-3}$, with $\phi_c=0.4$. The
grey band shows the noise amplitude $\sigma_0=C_0\sqrt{2\hhs e/(1+e)}$  over the
range of $e$ studied, which sets the scale of the fluctuations sustained in the
absence of structure; amplitudes far above it indicate genuine microphase separation.
Without noise, no structure forms at either quench depth within the simulated
time.
(c),(d)~Amplitude versus quench depth with the noise switched off ($C_0=0$),
starting either from a structured configuration (solid lines, filled symbols) or
from a homogeneous state (dashed lines, open symbols), for (c)~$\phi_c=0.4$ and
(d)~$\phi_c=0$. The shaded region marks the range $\Delta\tau\ge0$ over which
the two initial conditions relax to different steady states, i.e.\ where ordered
and homogeneous states coexist. Hysteresis is present for $\phi_c=0.4$, where the
effective free energy carries a cubic term, and absent for $\phi_c=0$, where it does not. Parameters are $\hhs=1$, $\hhl=10$, $\tau_4=0.4$, on a $128^3$ grid.}
\label{fig:A2}
\end{figure}

The two models differ more visibly in their response to noise just above the
spinodal. At $\Delta\tau=+0.01$ with $C_0=0.2$ (panel b), the full model develops
well-defined droplets whereas the simplified model shows only weak fluctuations.
Since both are the same distance above their respective spinodals, this
difference does not reflect the linear onset but the nonlinear terms, in which
the cubic contribution acts on $\phi_1+\phi_2$ rather than on $\phi_1$ alone. We examine the underlying mechanism in Fig.~\ref{fig:A2}.

\noindent\textbf{Effect of the exchange noise:} Figure~\ref{fig:A2} examines the role of the noise directly. Panels (a) and (b) show the RMS amplitude
$\langle\phi_1^2\rangle^{1/2}$ as a function of noise strength at $\Delta\tau=0$
and $\Delta\tau=0.01$, i.e., on and above the spinodal of the simplified model.
Without noise, no structure develops at either quench depth: at $\Delta\tau=0$ the
growth rate of the critical mode vanishes, thus an infinitesimal perturbation is
not amplified within the simulated time, and at $\Delta\tau>0$ the homogeneous
state is linearly stable. At $\Delta\tau=0$ the amplitude jumps at
$C_0=0.2$ to a plateau that is nearly independent of $e$ and lies a factor of
about five above $\sigma_0$, the scale of the fluctuations which the noise alone would
sustain in a structureless system; this indicates genuine microphase separation
rather than driven fluctuations, and structure therefore appears in a parameter
range where linear stability analysis predicts none. Above the spinodal, at
$\Delta\tau=0.01$, the amplitude remains at the fluctuation level up to
$C_0=0.3$ and rises only at $C_0=0.5$, and then only for the larger exchange
ratios, consistent with a nucleation barrier that decreases with $e$, since
$q_0^2$ and hence the cubic coefficient grow with $e$ while $\sigma_0$ does the
same. 

Panels (c) and (d) identify the mechanism. Switching the noise off and following
the system from either a structured or a homogeneous initial condition, we find
that for $\phi_c=0.4$ the structured state remains stable above the linear
instability: over a finite range $\Delta\tau\gtrsim0$ the two initial conditions
lead to different steady states, i.e., the ordered and homogeneous states coexist.
For $\phi_c=0$ they do not, and the structured state decays for all
$\Delta\tau\ge0$. Since $\phi_c$ is the only parameter that was varied between the two panels, and it enters the effective free energy through the cubic term
$-\tau_4\phi_c\tilde\phi_1^3$ of Eq.~\ref{eq:bmatrix},  we can attribute the coexistence to that term: the transition to the microphase is subcritical. This is consistent with the generic behavior of the Landau--Brazovskii model, the equilibrium counterpart at $\alpha=0$, which exhibits a first order transition in the presence of a cubic term in the free energy~\cite{brazovskiiphase1975}.

This provides the physical interpretation of the observed noise-induced structure formation. Because the transition
is subcritical, a finite-amplitude perturbation is required to reach the ordered
state, and the exchange noise supplies it, whereas the infinitesimal numerical
seed does not. The noise thus determines whether the ordered state is reached
within an accessible time, not whether it exists. We emphasize that this concerns
the approach to the ordered state, not the scaling laws reported in the main
text: the domain size selected below the spinodal is set by $q_0$ and is
insensitive to $C_0$ (Fig.~\ref{fig:A1}a), so the results of Sec.~IV. in the main text are unaffected. The one exception is the large-$q$ tail of $S(q)$, which
the noise lifts by adding a background $S(q)\simeq C_0^2\hhs e/[(1+e)q^4]$;
this is derived in Sec.~\ref{sec:noise}.

\section{Time evolution of Domain sizes}
\begin{figure}[h]
    \centering
    \includegraphics[width=0.8\linewidth]{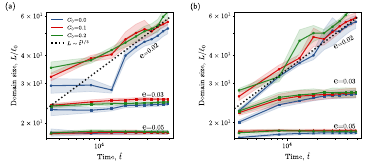}
    \caption{Time evolution of domain size in 3D phase separation with active noise. 
    Domain size $L/\ell_0$ as a function of time $\hht$ for different exchange ratios 
    ($e = 0.02, 0.03, 0.05$, represented by $\scalebox{1.2}{$\bullet$}$, $\scalebox{0.8}{$\blacksquare$}$, and $\blacktriangle$, respectively) 
    and noise amplitudes ($C_0 = 0.0, 0.1, 0.2$, shown in \textcolor{myblue}{blue}, \textcolor{myred}{red}, and \textcolor{mygreen}{green}, respectively). 
    Panels show results for quench depths (a)~$\Delta\tau = -0.01$ and (b)~$\Delta\tau = -0.05$. 
    System parameters: $L_x = L_y = L_z = 128\,\ell_0$. 
    Shaded regions indicate uncertainty from the fitting procedure. 
    The black dotted line represents the classical coarsening scaling law $L \sim \hht^{1/3}$, 
    characteristic of diffusion-limited domain growth.}
    \label{fig:domain_coarsening}
\end{figure}

Figure~\ref{fig:domain_coarsening} demonstrates how both the exchange ratio and noise amplitude influence coarsening dynamics, with higher noise levels generally leading to enhanced domain growth. To exclude early transient behavior, data analysis begins from $\hht \approx 3.3\times 10^3$. For $e < 0.03$, the domain size continues to grow during the whole simulation time following the classical scaling law $L \sim \hht^{1/3}$, characteristic of diffusion-limited coarsening. In contrast, for $e \geq 0.03$, domains reach a steady-state size. We provide four movies showing droplet evolution at $e=0.02$ and $e=0.05$ for quench depth $\Delta\tau=-0.05$ with and without noise ($C_0=0$ and $0.5$). Movies available in the online Supplemental Material.

\bibliographystyle{apsrev4-2}
\bibliography{Refs}